\documentclass{CHEP2006}

\usepackage{graphicx}
\usepackage{epsfig}

\begin{document}
\title{A configuration system for the ATLAS trigger}

\author{
H. von der Schmitt, MPI f\"ur Physik, Munich, Germany\\
N. Ellis, J. Haller, A. H\"ocker, T. Kohno, R. Spiwoks, CERN, Geneva, Switzerland \\
T. Wengler, University of Manchester, UK\\
A. dos Anjos, H. Zobernig, W. Wiedenmann, University of Wisconsin, Madison, USA \\
M. Landon, Queen Mary and Westfield College, London, UK
}

\maketitle

\begin{abstract}
The ATLAS detector at CERN's Large Hadron Collider will be exposed to 
proton-proton collisions from beams crossing at 40 MHz that have to be 
reduced to the few 100 Hz allowed by the storage systems. A three-level
trigger system has been designed to achieve this goal. We describe the
configuration system under construction for the ATLAS trigger chain.
It provides the trigger system with all the parameters 
required for decision taking and to record its history. The same system
configures the event reconstruction, Monte Carlo simulation and
data analysis, and provides tools for accessing and manipulating
the configuration data in all contexts.
\end{abstract}

\section{THE ATLAS TRIGGER}

The LHC proton bunches will cross at a frequency of approximately 40 MHz. 
The rate of events that can be committed to permanent storage in normal 
data taking is only a few 100\~Hz. The ATLAS trigger system faces the task 
to select the events that conform with the physics goals of ATLAS, 
among a dominant background of strong interaction processes.
The trigger system is organised in three levels. The first level trigger 
(LVL1)~\cite{LVL1TDR} utilises custom built hardware to derive a trigger 
decision within $2.5\mu{\rm s}$. The LVL1 decision is based on calorimeter 
information and on hits in the barrel and endcap muon trigger systems. 
The LVL1 systems deliver {\em Regions-of-Interest} (RoI) as {\em seeds} 
to the High Level Trigger (HLT) system ~\cite{HLTTDR}. The HLT consists of
two consecutive software triggers, Level-2 (LVL2) and Event Filter, which 
run on commodity PC farms.

At any point in time the complete trigger chain needs to be configured
in a consistent way. For LVL1, a trigger menu, comprising a
collection of event signatures that should cause a trigger, needs to be
defined and translated into a code the Central Trigger Processor (CTP) 
hardware can understand. Moreover the calorimeter and muon trigger
systems have to be configured such that they deliver the information
required by the trigger menu. The HLT starts from the RoIs delivered
by the LVL1 system and applies trigger decisions in a series of
steps, each refining existing information by acquiring additional data
from increasingly many sub-detectors. A list of physics signatures and
implemented event reconstruction (feature extraction) and selection 
algorithms is used to build signature and sequence tables for all HLT
trigger steps. The stepwise processing in the HLT is controlled by the 
{\em Steering}~\cite{STEERING}.

\begin{figure}[htb]
\centering
\epsfig{file=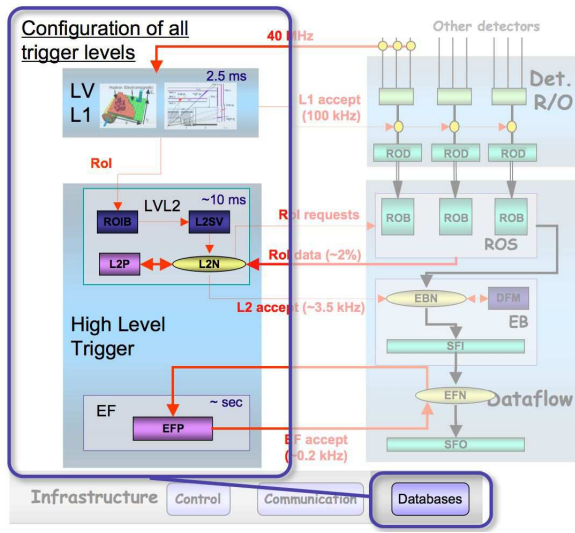,scale=0.42,angle=0., clip=0}
\caption{The trigger configuration schema in the context of the
	three-level trigger system (left) and the DAQ system (right) of the
	ATLAS experiment. }
\label{fig:context}
\end{figure}

The trigger configuration system has to comply with a number of complex 
use cases. When operating the experiment, the configuration parameters 
must be available to the systems participating in 
the LVL1 decision ({\em i.e.} the calorimeter trigger, the muon trigger, as well as the CTP) , and to all nodes forming the HLT farms. Figure~\ref{fig:context} 
depicts the trigger and its configuration system in the context of the 
ATLAS online framework.

Once a particular trigger configuration has been used in running, it
becomes essential history information for the data set obtained with it,
and must be remembered. Trigger configurations are expected to change 
frequently in response to varying experimental conditions. Tools must 
be provided to create new configurations and to guide the trigger expert 
by verifying their consistency. Furthermore the shift crews running the 
experiment need a tool to perform simple adjustments of the trigger
configuration throughout the lifetime of a beam coast.

In addition to data taking, the trigger configuration is an ingredient to 
data analysis and simulation. Users must be able to extract and use a 
trigger configuration in the context of the reconstruction, analysis and 
simulation software. This is required for trigger efficiency studies, 
trigger optimisation, and to determine the conditions of the data sets 
used in an analysis. In particular, trigger optimisation challenges the 
flexibility of the configuration system.

This paper describes the design of the trigger configuration system 
for the ATLAS experiment that meets the requirements outlined above.

\section{CONFIGURATION SYSTEM}

The trigger configuration system consists of a central relational database
(TriggerDB) that stores the configuration data, tools to populate the
database and ensure its consistency, and interfaces to extract the
stored data in preparation for data taking or other purposes 
({\em e.g.} simulation,
data analysis, etc.).  A schematic overview of the configuration system
is given in Fig.~\ref{fig:synopsis}.

\subsection{Trigger database}

The TriggerDB is the central part of the configuration system.  It is
used to store and protect all data that are needed to configure the
three levels of the ATLAS trigger: the LVL1 trigger menu, the HLT
trigger menu, the parameters ({\it job options}) of all software (SW) 
packages running in the HLT, and the corresponding release information.
Using the relational structure of the database, the various single data
elements ({\em e.g.} hardware registers of LVL1, algorithm parameters of HLT
selection algorithms or environment variables) are grouped
together to form bigger entities in a hierarchical tree-like
structure. Each element in the database is identified by a unique
integer {\em key}. These keys are used to construct larger entities higher
up in the hierarchy. The top-level entities, {\em i.e.} the ones
containing all information needed to configure all three trigger
levels are called {\it configurations}.  A configuration is composed
of one LVL1 configuration and one HLT configuration, which in turn
consist of other components like trigger menus and prescale sets
eventually leading to the basic configuration parameters\footnote
{
	As indicated in Fig.~\ref{fig:synopsis} the HLT configuration 
	can also be regarded as being composed of the HLT menu, the 
	algorithm parameters (HLT job options) and the HLT software 
	capabilities. The latter is used to enforce consistency 
	between the algorithms used in the configuration and the 
	capabilities of a SW release.
}.  
For the purpose of human readability all data entities are given a string 
name and a version number.  The tree-like structure described above allows 
one to reuse parts of a configuration when creating a new configuration,
by simply changing the referencing foreign keys in entities higher in 
the hierarchy, thereby avoiding unnecessary data duplication.

It is foreseen to store in the TriggerDB all versions of configurations 
that have been used for data taking and those prepared for simulation 
and test runs. The unique integer key (the {\em Master Key}) that identifies 
a certain configuration will be transfered to the {\em conditions database}
of the experiment~\cite{COOL}. This Master Key provides the unique
reference to a configuration and can therefore be used to retrieve the 
configuration at a later stage.

The TriggerDB is located on the same server as the conditions database
without, however, being embedded into its schema. Making use of the 
infrastructure provided by ATLAS and CERN-IT, the TriggerDB will follow 
all replication steps of the conditions database and will be available
at CERN and at external sites. The TriggerDB and all related tools are 
implemented to run on both MySQL and ORACLE.

It should be emphasised that the consistency of the configuration data
is an essential requirement that the configuration system must
fulfil. Inconsistent trigger configurations can lead to data loss or
data unusable for physics analysis. Wherever possible, the relational
schema has been designed to enforce consistency. Moreover, the
database population tools scrutinise the consistency of the data they
upload.

\subsection{Population Tools}

Due to the complexity of the trigger system and its configuration, the 
population of the database, including the composition of the trigger
menus, needs dedicated tools. At present two complementary systems 
are under development (see Fig.~\ref{fig:synopsis}):
\begin{enumerate}

\item	The {\em TriggerTool} is a stand-alone, java-based graphical user 
	interface to the TriggerDB. 

\item 	Custom {\em python scripts} convert the xml- and python-based 
	HLT menu and job configuration into SQL statements that 
	populate the database. The reverse mode where xml and job configuration 
	files are created from the database is also possible (see next Section).
	
\end{enumerate}

The TriggerTool is the central database population tool. It foresees
shift-crew and expert levels with different access
restrictions. Shifters can only choose among a list of approved
trigger menus and prescale sets to configure the {\em next
run}. Experts are allowed to modify existing and add new LVL1 and
HLT trigger menus. The TriggerTool handles the proper reordering of
the keys between the database tables. An important feature of the
TriggerTool is its capacity to perform automatic queries to validate
the validity of a trigger configuration. Examples for this are valid
collections of thresholds and prescale sets for LVL1, consistent
step-wise HLT signatures, and the coherent configurations of the HLT
feature extraction (event reconstruction) and selection algorithms,
each belonging to a unique software setup. The TriggerTool also
provides a convenient lightweight database browser for offline users,
providing advanced search functionality and access from remote
locations.

As indicated in Fig.~\ref{fig:synopsis}, the python scripts are used
to populate the HLT database tables. This includes the default configuration
properties of the HLT algorithms (for example feature extraction
options and selection requirements), and the dynamic link libraries,
services and tools required by the algorithms. These {\em components}
must be linked to the corresponding software release setups, which
requires that all the {\em capabilities} of the releases involved are
filled into the database (the capability of a release defines the available
features of the trigger software).  The database population is only
feasible by means of automatic release scanning tools, currently
implemented as python parsers.  The extracted information is written
to xml files, before being converted to SQL statements and uploaded to
the database. It is foreseen to perform such a scan for each new release,
identifying the changes between releases in the process. Specific
configuration of the algorithms, which goes beyond the default release
settings, must be inserted by hand into the database using the
TriggerTool.

Another ingredient needed is a compiler to translate the
human-readable LVL1 menu into the input files used to program the
look-up tables (LUT) and content addressable memory (CAM) that contain
the selection logic as part of the central trigger processor (CTP) of
the first level trigger. The compiler is implemented in C++ and can run in 
stand-alone mode taking the xml files extracted from the TriggerDB (see below). 
In addition, the compiler is integrated into the TriggerTool reading the
LVL1 menu from the TriggerDB. The output LUT and CAM files for each
LVL1 menu are then stored in the TriggerDB and made available for
online running.

\subsection{Data retrieval from the TriggerDB}
\label{sec:readout_interfaces}

\begin{figure}[t]
\begin{center}
\includegraphics*[width=85mm]{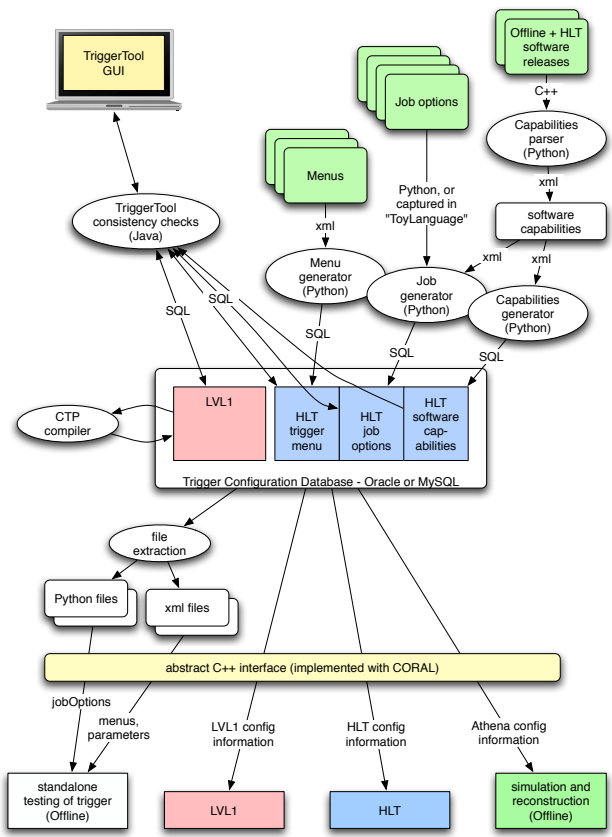}
\end{center}
\vspace{-0.5cm}
\caption{Schematic overview of the ATLAS trigger configuration system.
	Shown are the population methods of the trigger database
	(upper part) and the retrieval of configuration data from the database 
	for data taking (lower part). }
\label{fig:synopsis}
\end{figure}

There is a variety of use cases for data retrieval from the TriggerDB,
but the configuration of the complete system at the start of an online
data-taking run and the configuration of the offline simulation are
arguably the most challenging. Two independent data paths from the
TriggerDB are foreseen and have been implemented
(cf. Fig.~\ref{fig:synopsis}):

\begin{enumerate}

\item Configuration sets can be extracted from the TriggerDB into
  intermediate files (xml or python). These files can then be used by
  the user for stand-alone tests for, {\em e.g.}, development of new
  configurations and for tests of the online trigger system without
  interference with the TriggerDB during the commissioning of the
  system.
\item Configuration sets can be accessed by direct access to the
  TriggerDB. The various clients ({\em e.g.} LVL1 hardware modules or HLT
  processing nodes) contact the TriggerDB directly to get their
  configuration objects.
\end{enumerate}

To keep the differences between configuring via intermediate files and 
via direct database access at a minimum, both configuration paths 
make use of a common abstract interface. This interface is implemented in
C++ and is foreseen to be used online for data taking as well as for the
various offline use cases. Its two implementations are based on the Xerces 
xml parser  and the CORAL~\cite{CORAL} package allowing a vendor-independent 
access to the TriggerDB.

\section{DEPLOYMENT AND FIRST TESTS}

Tests of the LVL1 configuration system have been performed with the
offline simulation, yielding promising results. The two paths (via xml
or direct database access) can be routinely used to configure the simulation
of the LVL1 trigger. To complete the configuration system the C++
abstract interfaces need to be integrated with the online
state-machines controlling the various parts of the LVL1 system in
data-taking mode. As the number of clients in the LVL1 online system
is relatively small (about 20 CPUs controlling the hardware
crates of LVL1 and running the online state-machines) performance
problems are not expected for LVL1.

An initial test was performed of configuring the HLT from the
TriggerDB using a LVL2 muon selection chain. The complete
configuration of the muon chain together with the necessary auxiliary
services and tools were described in the TriggerDB. The muon selection
chain was run on a multi node system containing six LVL2 processing
units, a LVL2 supervisor and a read-out system (ROS)
emulator\footnote{See~\cite{HLTTDR} for a detailed description of the
HLT components.}. Events were pre-loaded to the ROS emulator and
retrieved by the processing units. Every LVL2 processing unit
retrieved the configuration from the TriggerDB. In a python module, the
retrieved information was converted in memory to standard python
configuration statements used for the ATLAS software framework Athena,
which were then used to set up all necessary Athena modules. In this
very early version it took about one second to retrieve all
configuration information from a single MySQL server. For the final system
it is envisaged to provide all necessary information in the TriggerDB
as a database view, which will then be read directly by a service
setting up the necessary software environment. This service is
presently under development.

Next steps in the development of the configuration system are the
integration of the system with the LVL1 hardware and the setup of more
complicated configurations for the HLT including more than one
selection chain. Performance studies and tuning will be an important
issue for the configuration of the large HLT processor farms in the
context of the online database architecture of the ATLAS experiment.

\end{document}